\documentclass[symmetry,article,submit,oneauthor]{mdpi}

\firstpage{1}
\pubvolume{xx}
\issuenum{1}
\articlenumber{5}
\pubyear{2020}
\copyrightyear{2020}
\history{Received: date; Accepted: date; Published: date}

\newcommand{\rmi}{\textrm{i}}
\newcommand{\rmd}{\textrm{d}}

\newcommand{\fl}{}

\Title{Supersymmetry of Relativistic Hamiltonians for Arbitray Spin}

\Author{Georg Junker $^{1,2}$\orcidA{}}

\AuthorNames{Georg Junker}

\address{%
$^{1}$ \quad European Southern Observatory, Karl-Schwarzschild-Stra{\ss}e 2, D-85748 Garching, Germany; gjunker@eso.org\\
$^{2}$ \quad Institut f\"ur Theoretische Physik I, Universit\"at Erlangen-N\"urnberg, Staudtstra{\ss}e 7, D-91058 Erlangen, Germany}

\corres{Correspondence: gjunker@eso.org}

\abstract{Hamiltonians describing the relativistic quantum dynamics of a particle with an arbitrary spin are shown to exhibit a supersymmetric structure when the even and odd elements of the Hamiltonian commute.
For such supersymmetric Hamiltonians an exact Foldy-Wouthuysen transformation exits which brings it into a block-diagonal form separating the positive and negative energy subspaces. Here the supercharges transform between energy eigenstates of positive and negative energy. The relativistic dynamics of a charged particle in a magnetic field is considered for the case of a scalar (spin-zero) boson obeying the Klein-Gordan equation, a Dirac (spin one-half) fermion and a vector (spin-one) boson characterised by the Proca equation.}

\keyword{Relativistic Wave Equation; Klein-Gordan Equation; Dirac Equation; Proca Equation; Supersymmetry}

\begin{document}
\section{Introduction}
Soon after the formulation of non-relativistic quantum mechanics by Heisenberg, Born, Jordan and Schr\"odinger in 1925 and 1926, Klein \cite{Klein1926} and Gordon \cite{Gordon1926} made first attempts to develop a relativistic quantum wave formalism. This Klein-Gordon equation is known to have certain deficits for its quantum mechanical interpretation but nowadays is well accepted as being the correct quantum wave formalism for spin-zero particles. To overcome the problems of the Klein-Gordan equation, Dirac in 1928 \cite{Dir1928a,Dir1928b} made an ansatz for a wave equation being linear in the time derivative and thus found his famous  equation describing the relativistic quantum dynamics of spin one-half fermions. The relativistic spin-one equation, also known as Proca equation, has been developed by Proca \cite{Proca1936} in 1936. In the same year Dirac \cite{Dir1936} and later Fierz and Pauli \cite{Fierz1939,FP1939} investigated relativistic wave equations for arbitrary spin. See also the later work by Bhabha \cite{Bhabha1945}. A group theoretical discussion of such wave equations was then given by Bargmann and Wigner in 1948 \cite{BW1948}.

In their fundamental work in 1950 Foldy and Wouthuysen \cite{FM1950} constructed a unitary transformation which separates the positive and negative energy states, that is, the Dirac Hamiltonian became block-diagonal. This work has triggered the study of the Hamiltonian form for the other wave equations. For example, Foldy \cite{Foldy1956} investigated the Klein-Gordan equation and Feshbach and Villars \cite{FV1958} made a unified approach to a Hamiltonian form of the Klein-Gordan and Dirac equation. The Hamiltonian form for the Proca equation has been studied by various authors including Duffin \cite{Duffin1938}, Kemmer \cite{Kemmer1939}, Yukawa, Sakata and Taketani \cite{YST1938,ST1940}, Corben and Schwinger \cite{CS1940}, as well as by Schr\"odinger \cite{Sch1954}. The problem of restoring a block-diagonal Hamiltonian via a so-called exact Foldy-Wouthuysen (FW) transformation for a particle with an arbitrary spin is still attracting many researchers. See, for example the recent work by Silenko \cite{Silenko2016} and Simulik \cite{Simulik2019} and references therein.

It is well-known \cite{Thaller92,Jun2019} that the Dirac Hamiltonian for a spin-$\frac{1}{2}$ particle in a magnetic field exhibits a $N=2$ supersymmetric (SUSY) structure and can be put into a block-diagonal form via an exact FW transformation. Here, the supercharges transform between positive and negative energy eigenstates and the eigenvalue problem of the Dirac Hamiltonian can be reduced to that for the non-relativistic Pauli Hamiltonian. The SUSY structure for the Klein-Gordon equation based on work by Feshbach and Villars \cite{FV1958} has briefly been discussed by Thaller in section 5.5.3 of his book \cite{Thaller92}. See also the more explicit discussion by Znojil \cite{Znojil2004}.

In the current paper we will generalise this formalism to relativistic Hamiltonians with arbitrary spin. For this we will recall in the next section the basic structure of such Hamiltonians. In section 3 we then show that whenever the odd part commutes with the even part of this Hamiltonian it is possible to construct a $N=2$ SUSY structure very similar to what is know for supersymmetric Dirac Hamiltonians \cite{Jun2019}. It is also shown that there exists an exact FW transformation bringing the Hamiltonian into a block-diagonal form. In section 4 we explicitly discuss the cases of a charged spin-zero, spin one-half and spin-one particle in an external magnetic field. In all three cases, which cover all the currently known charged elementary particles, we find that the eigenvalue problem of the relativistic Hamiltonian can indeed be reduced to that of a non-relativistic one. Section 5 discusses the resolvent of supersymmetric relativistic Hamiltonians and again shows that for the three cases under consideration the Green's function in essence may be reduce to that of the associated non-relativistic Hamiltonian. Finally in section 6 we present a short conclusion and an outlook for possible further investigations and in the appendix we collect some useful relations for the spin $s=1$ case which are not that commonly known.

\section{Relativistic Hamiltonians for arbitrary spin}
In the Hamiltonian form of relativistic quantum mechanics one puts the wave equation into a Schr\"odinger-like form
\begin{equation}\label{RelSE}
  \rmi\hbar \partial_t \Psi = H \Psi\,.
\end{equation}
The Hamiltonian in above equation in general is of the form
\begin{equation}\label{Hgen}
  H = \beta m +{\cal E} + {\cal O}\,,
\end{equation}
where $\beta^2 = 1$ acts as a grading operator and $m$ standard for the particle's mass. In addition to the mass term $\beta m$ the operator ${\cal E}$ represents the remaining even part of the Hamiltonian, that is, it commutes with the grading operator, $[\beta, {\cal E}] = 0$. The operator ${\cal O}$ denotes the odd part of $H$ and obeys the anticommutation relation $\{\beta , {\cal O}\}=0$. For a particle with spin $s$, $s=0,\frac{1}{2},1,\frac{3}{2},\ldots$, the Hilbert space ${\cal H}$ on which $H$ acts is given by
\begin{equation}\label{calH}
  {\cal H} = L^2(\mathbb{R}^3)\otimes\mathbb{C}^{2(2s+1)}\,,
\end{equation}
that is, the wave function $\Psi$ in (\ref{RelSE}) is a spinor with $2(2s+1)$ components \cite{Silenko2016}. Let us note that we can decompose the Hilbert space ${\cal H}$ into a direct sum of the two eigenspaces of the grading operator $\beta$ with eigenvalue $+1$ and $-1$, respectively,
\begin{equation}\label{calHpm}
  {\cal H} = {\cal H}_+\oplus {\cal H}_-\,, \qquad {\cal H}_\pm :=  L^2(\mathbb{R}^3)\otimes\mathbb{C}^{2s+1}\,.
\end{equation}
Obviously, ${\cal H}_\pm$ are simultaneously the subspaces where eigenvalues of $H$ are positive and negative, respectively.

For simplicity let us put the relativistic Hamiltonian (\ref{Hgen}) into the  form
\begin{equation}\label{Hanys}
  H=\beta{\cal M} + {\cal O}\,,
\end{equation}
where we have absorbed the mass $m$ in the even mass operator ${\cal M}:= m +\beta{\cal E}$ with $[\beta, {\cal M}] = 0$.
Let us note here that above Hamiltonian is self-adjoint, i.e.\ $H=H^\dag$, only for the case of fermions where $s=\frac{1}{2},\frac{3}{2},\ldots$ is a half-odd integer. For bosons, where $s$ takes integer values, the Hamiltonian is pseudo-hermitian, that is, $H=\beta H^\dag \beta$.

Choosing a representation where $\beta$ takes the diagonal form
\begin{equation}\label{beta}
  \beta = \left(\begin{array}{cc}
                     1 & 0 \\
                     0 & -1
                   \end{array}\right)\,,
\end{equation}
then the even and odd operators obeying $[\beta, {\cal M}] = 0$ and $\{\beta, {\cal O}\} = 0$ are necessarily of the form
\begin{equation}\label{MO}
  {\cal M} = \left(\begin{array}{cc}
                     M_+ & 0 \\ 0 & M_- \end{array}\right)\,,\qquad
  {\cal O} = \left(\begin{array}{cc}
                     0 & A \\ (-1)^{2s+1}A^\dag & 0 \end{array}\right)\,,
\end{equation}
where $M_\pm:{\cal H}_\pm\mapsto{\cal H}_\pm$ with $M^\dag_\pm= M_\pm$ is an operator mapping positive and negative energy states into positive and negative energy states, respectively. Whereas $A:{\cal H}_-\mapsto{\cal H}_+$ maps a negative energy state into a positive energy state and $A^\dag:{\cal H}_+\mapsto{\cal H}_-$ vice versa. With above representation the general relativistic spin-$s$ Hamiltonian then takes the form
\begin{equation}\label{H2}
  H = \left(\begin{array}{cc}
                      M_+& A \\ (-1)^{2s+1} A^\dag & -M_- \end{array}\right)\,.
\end{equation}

In the following section we will show that under the condition that the even and odd parts commute, i.e.\ $[{\cal M}, {\cal O}]=0$, the above Hamiltonian (\ref{H2}) exhibits a $N=2$ supersymmetric structure and allows for an exact Foldy-Wouthuysen transformation.

\section{Supersymmetric relativistic Hamiltonians for arbitrary spin}
Let us now assume that the even mass operator ${\cal M}$ and the odd operator ${\cal O}$ commute, that is, $[{\cal M}, {\cal O}]=0$. This condition implies that
\begin{equation}\label{MA}
  M_+ A = A M_-\,,\qquad  A^\dag M_+ = M_- A^\dag\,.
\end{equation}
As a consequence of this the squared Hamiltonian (\ref{H2}) becomes block diagonal
\begin{equation}\label{Hsqrt}
  H^2 = \left(\begin{array}{cc} M^2_+ + (-1)^{2s+1} AA^\dag& 0 \\ 0 & M^2_- +(-1)^{2s+1} A^\dag A \end{array}\right)\,.
\end{equation}
Inspired by the construction of a SUSY structure for supersymmetric Dirac Hamiltonians \cite{Jun2019} let us introduce the following SUSY Hamiltonian
\begin{equation}\label{HSUSY}
  H_{\rm SUSY} := \frac{(-1)^{2s+1}}{2mc^2}(H^2-{\cal M}^2) =
  \frac{1}{2mc^2} \left(\begin{array}{cc} AA^\dag& 0 \\ 0 & A^\dag A \end{array}\right)\geq 0\,
\end{equation}
and the complex supercharges
\begin{equation}\label{QQdag}
  Q:=\frac{1}{\sqrt{2mc^2}} \left(\begin{array}{cc} 0 & A \\ 0 & 0 \end{array}\right)\,,\qquad
  Q^\dag =\frac{1}{\sqrt{2mc^2}} \left(\begin{array}{cc} 0 & 0 \\ A^\dag & 0 \end{array}\right)\,.
\end{equation}
Here $m>0$ is an arbitrary mass-like parameter, representing, for example, the mass of the relativistic particle in (\ref{Hgen}).
It is obvious that these operators generate a transformation between positive and negative energy states.
A straightforward calculation shows that these operators together with the Witten parity operator $W:=\beta$ form a $N=2$ SUSY system, that is,
\begin{equation}\label{SUSYalgebra}
  H_{\rm SUSY} =\{Q,Q^\dag\}\,,\qquad \{Q,W\}=0\,,\qquad Q^2=0=(Q^\dag)^2\,,\qquad [W,H_{\rm SUSY}] = 0\,,\qquad W^2 = 1 \,.
\end{equation}
Let us note that ${\cal M}$ under condition (\ref{MA}) commutes with all operators of above algebra and thus constitutes a center of the SUSY algebra (\ref{SUSYalgebra}). Hence, a relativistic arbitrary-spin Hamiltonian (\ref{H2}) obeying the condition (\ref{MA}) may be called a {\it supersymmetric relativistic arbitrary-spin Hamiltonian}.

Let us also note that for a supersymmetric relativistic arbitrary-spin Hamiltonian there exists an exact Foldy-Wouthuysen transformation $U$ which brings (\ref{H2}) into the block-diagonal form \cite{Silenko2016}
\begin{equation}\label{HFW}
\fl
\begin{array}{rl}
  H_{\rm FW} := & U H U^\dag = \beta \sqrt{H^2} \\[2mm]
   = &\left(\begin{array}{cc} \sqrt{M^2_+ + (-1)^{2s+1} 2mc^2 H_+} & 0 \\ 0 & -\sqrt{M^2_- +(-1)^{2s+1} 2mc^2 H_-} \end{array}\right)\,,
\end{array}
\end{equation}
where the partner Hamiltonians $H_\pm$ are defined by
\begin{equation}\label{Hpm}
H_+ := \frac{1}{2mc^2}\,AA^\dag \,,\qquad H_- := \frac{1}{2mc^2}\, A^\dag A \,.
\end{equation}
In fact, it is known that under condition $[{\cal M}, {\cal O}]=0$ the exact Foldy-Wouthuysen transformation is explicitly given by \cite{Silenko2016,JunIno2018}
\begin{equation}\label{U}
  U:= \frac{|H|+\beta H}{\sqrt{2H^2+2{\cal M}|H|}}=\frac{1+\beta \,{\rm sgn}\, H}{\sqrt{2+\{{\rm sgn}\,H,\beta\}}}\,,\qquad {\rm sgn}\,H:=\frac{H}{\sqrt{H^2}}\,.
\end{equation}

As a side remark let us mention that the four projections operators
\begin{equation}\label{10a}
\textstyle
  P^\pm :=\frac{1}{2}\left[1\pm W \right]\,,\qquad \Lambda^\pm := \frac{1}{2}\left[1\pm {\rm sgn}\,H \right]\,,
\end{equation}
projecting onto the subspaces of positive/negative Witten parity and positive/negative eigenvalues of $H$, respectively, are related to each other via the same unitary transformation as $H$ and $H_{\rm FW}$
\begin{equation}\label{10b}
  P^\pm = U \Lambda^\pm U^\dag\,.
\end{equation}
That is, the positive and negative energy eigenspaces are transformed via $U$ into spaces of positive and negative Witten parity. In fact, one may verify that $U$ may be represented in terms of these projection operators as follows
\begin{equation}\label{UBL}
  U = \frac{P^+ \Lambda^+ +P^- \Lambda^-}{\sqrt{(P^+ \Lambda^+ + P^- \Lambda^- )( \Lambda^+ P^+ +\Lambda^- P^- )}}\,.
\end{equation}

The non-negative partner Hamiltonians $H_\pm\ge 0$ are essential isospectral which means that their strictly positive eigenvalues are identical. The corresponding eigenstates are related to each other via a SUSY transformation. To be more explicit let us assume these are given by
\begin{equation}\label{phi}
  H_\pm \phi^\pm_\varepsilon = \varepsilon\phi^\pm_\varepsilon \,, \qquad \phi^\pm_\varepsilon \in{\cal H}_\pm\,,\qquad \varepsilon >0\,,
\end{equation}
then the SUSY transformation reads \cite{Jun2019}
\begin{equation}\label{psiSUSY}
  \phi^+_\varepsilon = \frac{1}{\sqrt{2mc^2\varepsilon}}A\phi_\varepsilon^-\,,\qquad
  \phi^-_\varepsilon = \frac{1}{\sqrt{2mc^2\varepsilon}}A^\dag\phi_\varepsilon^+\,.
\end{equation}
Note that the energy eigenvalue $\varepsilon$ may be degenerate and above relations are valid for each of these energy eigenstates. We omit an additional index in $\phi^\pm_\varepsilon$ enumerating such a degeneracy.
In addition, both partner Hamiltonians $H_\pm$ may have a non-trivial kernel, that is, there may exist one or several eigenstates with
\begin{equation}\label{psi0}
  H_\pm \phi^\pm_0 = 0 \,.
\end{equation}
In this case SUSY is said to be unbroken \cite{Jun2019}. For these ground states, again we omit the index for a possible degeneracy, there exists no SUSY transformation relating $\phi^+_0$ and $\phi^-_0$. The breaking of SUSY can be studied via the so-called Witten index $\Delta$ \cite{Jun2019}, which in the current context is identical to the Fredholm index of $A$, if it is a Fredholm operator, that is,
\begin{equation}\label{WittenIndex}
  \Delta \equiv {\rm ind}\,A := {\rm dim\, ker}\,A - {\rm dim\, ker}\,A^\dag = {\rm dim\, ker}\,H_- - {\rm dim\, ker}\, H_+\,.
\end{equation}
Obviously a non-vanishing Witten index indicates that SUSY is unbroken. In connection with \cite{Thaller92}
\begin{equation}\label{DimKerQ}
  {\rm dim\, ker}\,(Q+Q^\dag)  = {\rm dim\, ker}\,A + {\rm dim\, ker}\,A^\dag = {\rm dim\, ker}\,H_- + {\rm dim\, ker}\, H_+
\end{equation}
the kernels of $H_\pm$, that is, the number of zero-energy states of $H_\pm$ are known. In general, however, the operator $A$ is not Fredholm and hence some regularized indices are studied \cite{Thaller92,Jun2019}.

Due to the SUSY condition (\ref{MA}) the mass operators commutes with the associated partner Hamiltonians,  $[M_\pm , H_\pm ] = 0$, and therefore have an identical set of eigenstates. Let us denote the corresponding eigenvalues of $M_\pm$ by $m_\pm c^2$, that is
\begin{equation}\label{psiM}
  M_\pm\phi^\pm_\varepsilon = m_\pm c^2\phi^\pm_\varepsilon\,,\qquad \varepsilon\ge 0\,,
\end{equation}
then obviously the eigenvalues and eigenstates of (\ref{HFW}),
\begin{equation}\label{HFWPsi}
  H_{\rm FW}\psi^\pm_\varepsilon = E_\pm \psi^\pm_\varepsilon\,,
\end{equation}
are given by
\begin{equation}\label{EpmPsipm}
\fl
  E_\pm=\pm\sqrt{m^2_\pm c^4+(-1)^{2s+1}2mc^2\varepsilon}\,,\qquad
  \psi^+_\varepsilon = \left(\begin{array}{c}\phi^+_\varepsilon \\ 0 \end{array}\right)\,,\qquad
  \psi^-_\varepsilon = \left(\begin{array}{c}0 \\ \phi^-_\varepsilon \end{array}\right)\,.
\end{equation}
Here let us note that the mass eigenvalues may depend on the energy eigenvalues, $m_\pm=m_\pm(\varepsilon)$. Using the relation (\ref{MA}) in combination with the SUSY relation (\ref{psiSUSY}) and the above eigenvalue equation (\ref{psiM}) one may verify that $m_+(\varepsilon) = m_-(\varepsilon)$ for all $\varepsilon >0$. In essence this means that the spectrum of a supersymmetric relativistic Hamiltonian is symmetric about zero with a possible exception at $\pm m_\pm(0) c^2$, that is if $\varepsilon = 0$, which may only occur in the case of unbroken SUSY.

The eigenstates of the original Hamiltonian (\ref{H2}) are then easily found via the unitary transformation $\Psi^\pm_\varepsilon = U^\dag \psi^\pm_\varepsilon $ having the same eigenvalues $E_\pm$ given above.
Hence, the eigenvalue problem of a supersymmetric relativistic spin-$s$ Hamiltonian can be reduced to the simultaneous eigen\-value problems for $M_\pm$ and $H_\pm$ on ${\cal H}_\pm$.

It will turn out in the examples to be discussed below that the partner Hamiltonians $H_\pm$ and the mass operators $M_\pm$ are in essence represented by a non-relativistic Schr\"odinger-like Hamiltonian $H_{\rm NR}$ and/or some constant operator. To be more precise we will show for all three case $s=0,\frac{1}{2}$ and $1$ discussed below that the FW-transformed relativistic Hamiltonian takes the form
\begin{equation}\label{HFWgen}
  H_{\rm FW}=\beta mc^2 \sqrt{1+ \frac{2H_{\rm NR}}{mc^2}}
\end{equation}
with $H_{\rm NR}$ representing the associate non-relativistic Hamiltonian as
\begin{equation}\label{HNRlimit}
  \lim_{c\to\infty}\left( P^\pm H_{\rm FW} P^\pm \mp mc^2 \right)=\pm\lim_{c\to\infty}\left( mc^2\sqrt{1+2H_{\rm NR}/mc^2} - mc^2 \right) = \pm H_{\rm NR}\,.
\end{equation}
\section{Examples}
In below subsections we will consider a relativistic charged particle with charge $e$ and mass $m$ in an external magnetic field $\vec{B}:=\vec{\nabla}\times\vec{A}$ characterised by a vector potential $\vec{A}$. The symbol $\vec{\pi}$ stands for the kinetic momentum, that is, $\vec{\pi}:= \vec{p}-e\vec{A}/c$, where $c$ denotes the speed of light.
We will consider the case of a scalar particle with spin $s=0$, a Dirac particle with $s=\frac{1}{2}$ and a vector boson having spin $s=1$.

\subsection{The Klein-Gordon Hamiltonian with magnetic field}
The Schr\"odinger form of the Klein-Gordon equation as been considered by Feshbach and Villars in \cite{FV1958} where they have shown that the Klein-Gordon equation for a charged particle in a magnetic field can be put into the Schr\"odinger form (\ref{RelSE}) where the pseudo-Hermitian Hamiltonian is given by
\begin{equation}\label{HKG}
  H=\left(\begin{array}{cc} mc^2 + \frac{\vec{\pi}^2}{2m} & \frac{\vec{\pi}^2}{2m} \\[2mm] -\frac{\vec{\pi}^2}{2m} & -mc^2 -\frac{\vec{\pi}^2}{2m} \end{array}\right)\,,\qquad {\cal H}=L^2(\mathbb{R}^3)\otimes\mathbb{C}^2\,.
\end{equation}
Comparing this with the general form (\ref{H2}) we may identify the operators $M_\pm$ and $A$ as follows.
\begin{equation}\label{KGMA}
  M_\pm =\frac{1}{2m}\vec{\pi}^2+mc^2\,,\qquad A = \frac{1}{2m}\vec{\pi}^2\,.
\end{equation}
Obviously these operators are identical up to an additional constant $M_\pm = A +mc^2$ and hence the condition (\ref{MA}), that is $[M_\pm,A]=0$ is trivially fulfilled.  In other words the Klein-Gordon Hamiltonian (\ref{HKG}) for a charged scalar particle in the presence of an arbitrary magnetic field represents a supersymmetric relativistic spin-zero Hamiltonian. Let us note that the two operators (\ref{KGMA}) in essence are given by the Landau Hamiltonian
$H_{\rm L}:= (\vec{p}-e\vec{A}/c)^2/2m=\vec{\pi}^2/2m$ of a non-relativistic spinless charged particle of mass $m$ in a magnetic field, that is, $M_\pm = H_{\rm L} +mc^2$ and $A= H_{\rm L}$. As a consequence the eigenvalue problem for (\ref{HKG}) is reduced to that of $H_{\rm L}$. The FW transformed Hamiltonian explicitly reads
\begin{equation}\label{HFWKG}
\fl
  H_{\rm FW}=\left(\begin{array}{cc} \sqrt{(H_{\rm L} + mc^2)^2 -H_{\rm L}^2} & 0 \\ 0 & -\sqrt{(H_{\rm L} + mc^2)^2 -H_{\rm L}^2} \end{array}\right)
  =\beta mc^2 \sqrt{1+ \frac{2H_{\rm L}}{mc^2}}\,.
\end{equation}
Let us denote the eigenvalues of $H_{\rm L}$ by $\epsilon$ then the eigenvalues of $M_\pm$ and $H_\pm =H_{\rm L}^2/2mc^2$ read
\begin{equation}\label{KGEV}
  m_\pm c^2=mc^2+\epsilon\,,\qquad \varepsilon = \frac{\epsilon^2}{2mc^2}
\end{equation}
and via relation (\ref{EpmPsipm}) we find the eigenvalues of the Klein-Gordon Hamilonian
\begin{equation}\label{EKG}
\fl
  E_\pm=\pm\sqrt{m_\pm^2c^4-2mc^2\varepsilon}=\pm\sqrt{(mc^2+\epsilon)^2-\epsilon^2}=\pm mc^2\sqrt{1+\frac{2\epsilon}{mc^2}}\,,
\end{equation}
which is a result expected from relation (\ref{HFWKG}).
For a non-vanishing constant magnetic field, say in $z$-direction $\vec{B}=B\vec{e}_z$ with $B\neq 0$, the eigenvalues of $H_{\rm L}$ are the well-known Landau levels \cite{Fock1928,Landau1930}
\begin{equation}\label{LL}
\fl
  \epsilon = \hbar\omega_c \left(n+\frac{1}{2}\right) + \frac{\hbar^2k_z^2}{2m}\,,\qquad n\in\mathbb{N}_0\,,\qquad k_z\in\mathbb{R}\,,\qquad \omega_c:=\frac{|eB|}{mc}\,.
\end{equation}
As $\epsilon \geq \hbar\omega_c/2 > 0$ so is $\varepsilon > 0$ and hence ${\rm dim\, ker}\,H_- = {\rm dim\, ker}\, H_- = 0$. In other words, the Witten index (\ref{WittenIndex}) vanishes and SUSY is broken for the Klein-Gordon Hamiltonian in a constant magnetic field.

\subsection{The Dirac Hamiltonian with magnetic field}
The Dirac equation representing the relativistic dynamics of spin-$\frac{1}{2}$ fermions has intensively been studied since its introduction. See for example the excellent book by Thaller \cite{Thaller92}. For a charged particle in an arbitrary external magnetic field the Dirac Hamiltonian reads
\begin{equation}\label{HDE}
  H=\left(\begin{array}{cc} mc^2 & c\vec{\sigma}\cdot \vec{\pi} \\  c\vec{\sigma}\cdot \vec{\pi} & -mc^2 \end{array}\right)\,,
\end{equation}
where $\vec{\sigma}=(\sigma_1,\sigma_2,\sigma_3)^T$ stands for a three-dimensional vector who's components are given by the Pauli matrices acting on $\mathbb{C}^2$, thus representing the spin-$\frac{1}{2}$ degree of freedom. Comparing this with the general form (\ref{H2}) we may identify the operators $M_\pm = mc^2$ and $A=c\sigma\cdot \vec{\pi}=A^\dag$ and note that condition (\ref{MA}) is trivially fulfilled. Hence, the Dirac Hamiltonian (\ref{HDE}) is indeed supersymmetric and its FW transformed form is know \cite{JunIno2018} to be expressible in terms of the non-relativistic
Pauli Hamiltonian for a spin-$\frac{1}{2}$ particle with Land\'{e} $g$-factor $g=2$.
\begin{equation}\label{HPE}
  H_{\rm P}:= \frac{1}{2m}(\vec{\sigma}\cdot\vec{\pi})^2 = \frac{1}{2m}(\vec{p}-e\vec{A}/c)^2-\frac{e\hbar}{mc}\vec{B}\cdot\vec{\sigma}\,.
 \end{equation}
Obviously the partner Hamiltonians can be identified with the Pauli Hamiltonian $H_\pm = A^2/2mc^2= H_{\rm P}$ and we find for the FW  transformed Dirac Hamiltonian
\begin{equation}\label{HFWD}
\fl
  H_{\rm FW}=\left(\begin{array}{cc} \sqrt{m^2c^4 + 2mc^2 H_{\rm p}} & 0 \\ 0 & -\sqrt{m^2c^4 + 2mc^2 H_{\rm p}} \end{array}\right)
  =\beta mc^2 \sqrt{1+ \frac{2H_{\rm P}}{mc^2}}\,.
\end{equation}
As shown by Avron and Casher \cite{AC78}, for a magnetic field having only a $z$-component, $\vec{B}=B(x,y)\vec{e}_z$,  the ground-state energy of $H_{\rm P}$ is zero and the Witten index (\ref{WittenIndex}) in essence is given by the flux $F=\int_{\mathbb{R}^2}\rmd x \rmd y B(x,y)$ measured in units of the magnetic flux quantum $\Phi_0:= 2\pi\hbar c/|e|$. Hence SUSY is unbroken in this case. For details, see for example \cite{Jun2019}, where it is shown that the Pauli Hamiltonian for this magnetic field provides an additional SUSY structure within both subspaces ${\cal H}_\pm$.

Finally, for an electron ($e<0$) in a constant magnetic field the eigenvalues of $H_{\rm P}$ are well-known
\begin{equation}\label{HPeps}
\fl
  \varepsilon = \hbar\omega_c \left(n+\frac{1}{2}+s_z\right) + \frac{\hbar^2k_z^2}{2m}\,,\qquad n\in\mathbb{N}_0\,,\qquad k_z\in\mathbb{R}\,,\qquad \textstyle s_z\in\{-\frac{1}{2},\frac{1}{2}\}\,,
\end{equation}
and the eigenvalues for (\ref{HDE}) determined via (\ref{EpmPsipm}) are the relativistic Landau levels first obtained by Rabi in 1928 \cite{Rabi1928}
\begin{equation}\label{EDH}
  E_\pm=\pm\sqrt{m^2c^2+\hbar^2c^2k_z^2 +2mc^2\hbar\omega_c(n+1/2+s_z)}\,.
\end{equation}
The degeneracy for each set of quantum numbers $(n,s_z,k_z)$ is given by the largest integer which is stricly less than $|F|/\Phi_0$ and is only finite in case the magnetic field has a compact support.
\subsection{The Spin-1 Hamiltonian with magnetic field}
Initiated by Proca's work \cite{Proca1936} several authors have studied relativistic spin-one wave equation. Let us mention here the work by Duffin \cite{Duffin1938}, by Kemmer \cite{Kemmer1939} and by Yukawa, Sakata and Taketani \cite{YST1938,ST1940}. An early study of the  eigenvalue problem is due to Corben and Schwinger \cite{CS1940}. A relativistic Hamiltonian was studied, for example, by
Young and Bludman \cite{YB1963}, by Krase et al \cite{KLG1971} and by Tsai and coworkers \cite{TY1971,GT1971}. The later work by Daicic and Frankel \cite{DF1993} presents an alternative solution to eigenvalue problem of the spin-one Hamiltonian in an external magnetic field. A recent treatment via FW transformation can be found in ref.\ \cite{Sil2014}.

The Hamiltonian of a charged spin-one particle with a priori arbitrary $g$-factor is given by, see for example \cite{YB1963,DF1993,Sil2014},
\begin{equation}\label{HPrE}
\fl
  H=\left(\begin{array}{cc}
    mc^2 + \frac{\vec{\pi}^2}{2m}-\frac{ge\hbar}{2mc}(\vec{S}\cdot\vec{B}) &
                       \frac{\vec{\pi}^2}{2m}-\frac{1}{m}(\vec{S}\cdot\vec{\pi})^2+\frac{(g-2)e\hbar}{2mc}(\vec{S}\cdot\vec{B}) \\[2mm]
    -\frac{\vec{\pi}^2}{2m}+\frac{1}{m}(\vec{S}\cdot\vec{\pi})^2-\frac{(g-2)e\hbar}{2mc}(\vec{S}\cdot\vec{B})  &
                       -mc^2 -\frac{\vec{\pi}^2}{2m} +\frac{ge\hbar}{2mc}(\vec{S}\cdot\vec{B}) \end{array}\right)\,,
\end{equation}
where $\vec{S}=(S_1,S_2,S_3)^T$ is a vector who's components are $3\times 3$ matrices acting on $\mathbb{C}^3$ and obeying the $SO(3)$ algebra $[S_i,S_j]=\rmi \varepsilon_{ijk}S_k$ representing the spin-one-degree of freedom of the particle. Again we may identify the operators
\begin{equation}\label{PEMA}
\fl
  M_\pm := mc^2 + \frac{\vec{\pi}^2}{2m}-\frac{ge\hbar}{2mc}(\vec{S}\cdot\vec{B}) \,,\qquad
  A := \frac{\vec{\pi}^2}{2m}-\frac{1}{m}(\vec{S}\cdot\vec{\pi})^2+\frac{(g-2)e\hbar}{2mc}(\vec{S}\cdot\vec{B}) =A^\dag\,.
\end{equation}
From now on let us assume that the magnetic field $\vec{B}$ is constant, i.e.\ $\vec{A}=\frac{1}{2}\vec{B}\times\vec{r}$. Under this condition one may verify that
\begin{equation}\label{AMS=1}
  [M_\pm,A]=(g-2)\frac{e\hbar}{2m^2c}\left[(\vec{S}\cdot\vec{B}),(\vec{S}\cdot\vec{\pi})^2 \right]\,.
\end{equation}
Hence, the condition (\ref{MA}) is fulfilled if $g=2$. In other words the relativistic spin-one Hamiltonian (\ref{HPrE}) is a supersymmetric Hamiltonian if the gyromagnetic factor is given by $g=2$. For a detail discussion we refer to the recent paper \cite{Jun2020}.
Here we remark that the "Vector Boson" Hamiltonian
\begin{equation}\label{H0}
  H_{\rm V} := \frac{\vec{\pi}^2}{2m}-\frac{e\hbar}{mc}(\vec{S}\cdot\vec{B})
\end{equation}
represents the non-relativistic Hamiltonian of a charged spin-one particle in a magnetic field with gyromagnetic factor $g=2$.
Note that (\ref{H0}) is related to the quantity $\alpha$ introduced by Weaver \cite{Weaver1976} by $H_V=\frac{\alpha}{2m}$.
That this is indeed the non-relativistic version of (\ref{HPrE}) was already mention in ref.\ \cite{KLG1971}.
With this we have $M_\pm = H_{\rm V}+mc^2$ and with relation (3.18) and (3.19) from ref.\ \cite{DF1993}, see also the appendix A, a straightforward calculation shows that $A^2=H_{\rm V}^2$. That is, the partner Hamiltonians read $H_\pm = H_{\rm V}^2/2mc^2$ and the transformed FW Hamiltonian takes the form
\begin{equation}\label{HFWPE}
\fl
  H_{\rm FW}=\left(\begin{array}{cc} \sqrt{(H_{\rm V} + mc^2)^2 -H_{\rm V}^2} & 0 \\ 0 & -\sqrt{(H_{\rm V} + mc^2)^2 -H_{\rm V}^2} \end{array}\right)
  =\beta mc^2 \sqrt{1+ \frac{2H_{\rm V}}{mc^2}}
\end{equation}
The eigenvalues of (\ref{H0}) are given by (we assume $e<0$)
\begin{equation}\label{HVeps}
\fl
  \epsilon = \hbar\omega_c \left(n+\frac{1}{2}+s_z\right) + \frac{\hbar^2k_z^2}{2m}\,,\qquad n\in\mathbb{N}_0\,,\qquad k_z\in\mathbb{R}\,,\qquad \textstyle s_z\in\{-1,0,1\}\,.
\end{equation}
Hence the spectrum of the partner Hamiltonians $H_\pm$ is given by $\varepsilon=\epsilon^2/2mc^2$
and SUSY is unbroken as $\epsilon = 0$ for $n=0$, $s_z = -1$ and $k_z = \pm 1/\lambda_L$ with $\lambda_L := \sqrt{\hbar/m\omega_c}=\sqrt{\hbar c/|eB|}$ being the Lamor length \cite{DF1993}. That is, SUSY is unbroken for a spin-1 particle in a homogeneous magnetic field but the Witten index remains zero as $H_+=H_-$ and therefore $\dim {\rm ker}\, H_+ = \dim  {\rm ker}\, H_-$ .
The corresponding eigenvalues of (\ref{HPrE}) are given by
\begin{equation}\label{EPrH}
  E_\pm=\pm\sqrt{m^2c^2+\hbar^2c^2k_z^2 +2mc^2\hbar\omega_c(n+1/2+s_z)}\,,
\end{equation}
which is identical in form to the Dirac case (\ref{EDH}) but $s_z$ now taking the integer values as given in (\ref{HVeps}). In fact, for $k_z=0$, $n=0$ and $s_z=-1$ above eigenvalues would become complex if $|B|>m^2c^3/|e|\hbar$. This limit would imply $\lambda_L < \lambda_C := \hbar/mc$, that is, the Lamor wavelength being smaller than the Compton wavelength of the vector boson. Note that confining a quantum particle to a region of the order of its Compton wavelength $\Delta x \sim \lambda_C$ implies by the uncertainty relation a momentum fluctuation $\Delta p \sim mc$ and thus a single particle description is no longer appropriate. In other words for such large magnetic fields a description via quantum field theory must be applied.
\section{The resolvent of supersymmetric relativistic arbitrary-spin Hamiltonians}
In this section we want to study the resolvent or Green's function of  supersymmetric relativistic arbitrary-spin Hamiltonians defined as
\begin{equation}\label{G}
  G(z) := \frac{1}{H-z}\,,\qquad z\in\mathbb{C}\backslash {\rm spec}\, H\,.
\end{equation}
For this is it convenient to first look at the iterated resolvent which is given by
\begin{equation}\label{g}
  g(\zeta) := \frac{1}{H^2-\zeta}\,,\qquad \zeta\in\mathbb{C}\backslash {\rm spec}\, H^2
\end{equation}
and is related with (\ref{G}) via the obvious relation
\begin{equation}\label{G2}
  G(z) = (H+z)\,g(z^2)\,.
\end{equation}
As $H^2$ is block-diagonal so is $g$ and hence it can be put into the form
\begin{equation}\label{g_pm}
  g(\zeta) := \left(\begin{array}{cc} g^+(\zeta) & 0 \\ 0 &  g^-(\zeta)\end{array}\right)
  \qquad \mbox{with} \qquad
  g^\pm(\zeta) := \frac{1}{M_\pm^2+ (-1)^{2s+1}2mc^2 H_\pm - \zeta}\,.
\end{equation}
As a result the resolvent (\ref{G}) can be expressed in terms of (\ref{g_pm}) as follows.
\begin{equation}\label{G3}
  G(z) = \left(\begin{array}{cc} (z+M_+ )g^+(z^2) & A g^-(z^2)\\[2mm] (-1)^{2s+1}A^\dag g^+(z^2)&  (z-M_-)g^-(z^2)\end{array}\right).
\end{equation}
In the following subsections we will now explicitly consider the three cases discussed in the previous section. It will turn out that for these three cases the diagonal elements $g^\pm$ of the iterated Green's function can be expressed in terms of the Green's function of the corresponding non-relativistic Hamiltonian $H_{\rm NR}$, that is,
\begin{equation}\label{GLgen}
g^\pm(\zeta) = \frac{1}{2mc^2}\,G_{\rm NR}\left( \textstyle \frac{\zeta}{2mc^2}-\frac{mc^2}{2} \right)\,,\qquad
G_{\rm NR}(\xi) := \frac{1}{H_{\rm NR} - \xi}\,,\qquad\xi\in\mathbb{C}\backslash {\rm spec}\,H_{\rm NR}.
\end{equation}
Note that the relation $\xi = z^2/2mc^2 - mc^2/2$, which can be put into the form $z=\pm mc^2\sqrt{1+2\xi /mc^2}$, in essence reflects the relation (\ref{HFWgen}).

\subsection{The resolvent of the Klein-Gordon Hamiltonian with magnetic field}
Following the discussion of the first example of section 4 we may express all relevant operators in terms of the Landau Hamiltonian $H_{\rm L}=\vec{\pi}^2/2m$. Explicitly we have
\begin{equation}\label{KGOps}
  M_\pm = H_{\rm L} + mc^2\,,\qquad
  H_\pm = H_{\rm L}^2 / 2mc^2\,,\qquad
  A = H_{\rm L} \dag\,,
\end{equation}
which results in the iterated resolvents
\begin{equation}\label{g_pm0}
  g^\pm(\zeta) := \frac{1}{(H_{\rm L}+mc^2)^2-H^2_{\rm L} - \zeta}=\frac{1}{2mc^2}G_{\rm L}\left(\textstyle \frac{\zeta}{2mc^2}-\frac{mc^2}{2}\right)\,,
\end{equation}
where $G_{\rm L}$ stands for the Green function of the Landau Hamiltonian in terms of which the Klein- Gordan Hamiltonian reads
\begin{equation}\label{HKG_HL}
H=\left(\begin{array}{cc}mc^2+H_{\rm L} & H_{\rm L} \\ -H_{\rm L}& -mc^2-H_{\rm L} \end{array} \right) \,.
\end{equation}
The Green's function then reads in terms of the Landau Hamiltonian
\begin{equation}\label{Gs=0}
  G(z) = \frac{1}{2mc^2}
         \left( \begin{array}{cc} z+mc^2+H_{\rm L} & H_{\rm L} \\ -H_{\rm L} & z-mc^2-H_{\rm L} \end{array} \right)
         G_{\rm L}\left(\textstyle \frac{z^2}{2mc^2}-\frac{mc^2}{2}\right)\,.
\end{equation}
\subsection{The resolvent of the Dirac particle in a magnetic field}
As in the above discussion let us first recall the observations made in section 4.2, that is,
\begin{equation}\label{DOps}
  M_\pm = mc^2\,,\qquad
  H_\pm = A^2 / 2mc^2 = H_{\rm P}\,,\qquad
  A = c\vec{\sigma}\cdot\vec{\pi}\,,
\end{equation}
which provide us with the components of the iterated kernel
\begin{equation}\label{g_pm1/2}
  g^\pm(\zeta) := \frac{1}{2mc^2H_{\rm P}+m^2c^4 - \zeta}=\frac{1}{2mc^2}G_{\rm P}\left(\textstyle \frac{\zeta}{2mc^2}-\frac{mc^2}{2}\right)\,,
\end{equation}
where $G_{\rm P} (\epsilon) := \left(H_{\rm P} - \epsilon\right)^{-1}$ is the resolvent of the non-relativistic Pauli Hamiltonian. In terms of this Pauli Green's function and the spin projection operator $A$ the Dirac Green's function can be put into the form
\begin{equation}\label{Gs=1/2}
  G(z) = \frac{1}{2mc^2}
         \left( \begin{array}{cc} z+mc^2 & A \\ A & z-mc^2 \end{array} \right)
         G_{\rm P}\left(\textstyle \frac{z^2}{2mc^2}-\frac{mc^2}{2}\right)\,.
\end{equation}
Some explicit examples have been worked out in ref.\ \cite{JunIno2018}.
\subsection{The resolvent of a vector boson in a magnetic field}
From subsection 4.3 let us recall the relevant operators as follows
\begin{equation}\label{VOps}
  M_\pm = H_{\rm V} +mc^2\,,\qquad
  H_\pm = H_{\rm V}^2 / 2mc^2 \,,\qquad
  A = \frac{\vec{\pi}^2}{2m}-\frac{1}{m}(\vec{S}\cdot\vec{\pi})^2\,,
\end{equation}
where the vector Hamilton $H_{\rm V}$ is given in eq.\ (\ref{H0}). Recalling that $A^2=H_{\rm V}^2 $ we find for the iterated Green's functions
\begin{equation}\label{g_pm1}
  g^\pm(\zeta) :=  \frac{1}{(H_{\rm V}+mc^2)^2-H^2_{\rm V} - \zeta}=\frac{1}{2mc^2}G_{\rm V}\left(\textstyle \frac{\zeta}{2mc^2}-\frac{mc^2}{2}\right)
\end{equation}
with $G_{\rm V} (\epsilon) := \left(H_{\rm V} - \epsilon\right)^{-1}$. The relativistic spin-one Hamiltonian explicitly reads
\begin{equation}\label{HKG_HV}
H=\left(\begin{array}{cc}mc^2+H_{\rm V} & A \\ -A & -mc^2-H_{\rm V} \end{array} \right) \,
\end{equation}
and leads us to the Green's function
\begin{equation}\label{Gs=1}
  G(z) = \frac{1}{2mc^2}
         \left( \begin{array}{cc} z+mc^2+H_{\rm V} & A \\ -A & z-mc^2-H_{\rm V} \end{array} \right)
         G_{\rm V}\left(\textstyle \frac{z^2}{2mc^2}-\frac{mc^2}{2}\right)\,.
\end{equation}


\section{Summary and Outlook}
In this work we have considered relativistic one-particle Hamiltonians for an arbitrary but fix spin $s$ and have shown that under the condition, that its even part commutes with its odd part, a SUSY structure can be established. Here the SUSY transformations map states of negative energy to those of positive energy and vice versa. This is different to the usual SUSY concepts in quantum field theory where those charges transform bosonic into fermionic states and vice versa. As examples we have chosen the physically most relevant cases of a massive charged particle in a magnetic field for the cases of a scalar particle ($s=0$), a Dirac fermion ($s=1/2$) and a vector boson ($s=1$). In the case of a constant magnetic field SUSY is broken for $s=0$ but remains unbroken for $s=1/2$ and $s=1$. The Witten index is only non-zero in the Dirac case but vanishes for the bosonic cases discussed. However, all three cases have resulted in the notable observation (\ref{HFWgen}) that the FW-transformed Hamiltonian $H_{\rm FW}$ is entirely expressible in terms of a corresponding non-relativistic Hamiltonian $H_{\rm NR}$. As $H^2_{\rm FW}=H^2$ the relativistic energy-momentum relation can be put into the form
\begin{equation}\label{relEPrelation}
  H^2=m^2c^4+2mc^2H_{\rm NR}\,,
\end{equation}
which allows us to related $H_{\rm NR}$ with the SUSY Hamiltonian (\ref{HSUSY}).

There naturally arises the desire to also study the higher-spin cases $s\ge 3/2$. The corresponding free-particle Hamiltonians have been constructed, for example, by Guertin \cite{Guertin1974} in a unified way. However, as Guertin mentions, only for the cases discussed here, i.e.\ $s=0,1/2$ and $1$, the corresponding Hamiltonians are local operators.

Another route for further investigation would be to consider more exotic magnetic fields. For example, choosing an imaginary vector potential such that the kinetic momentum takes the form $\vec{\pi}=\vec{p}+\rmi m\omega\vec{r}$ in essence leads for $s=1/2$ to the so-called Dirac oscillator, which is know to exhibit such a SUSY structure \cite{JunIno2018}. To the best of our knowledge the corresponding Klein-Gordon and vector boson oscillators have not yet been studied in the context of SUSY. Similarly, following the discussion of ref.\ \cite{JunIno2018} on the Dirac case, one may extend these discussion on a path-integral representation of the iterated Green's functions to the bosonic cases $s=0$ and $s=1$.

\acknowledgments{The author has enjoyed enlightening discussions with Simone Warzel for which I am very thankful.}

\conflictofinterest{The author declares no conflicts of interest.}

\appendix
\section{Some useful relations for the spin-one case}
In this appendix we present a few relations which provide some additional steps used in section 4.3. For an arbitrary magnetic field let us recall that the components of the kinetic momentum given by $\pi_j=p_j-(e/c)A_j$ obey the commutation relation
\begin{equation}\label{picomute}
  [\pi_k,\pi_l] =( \rmi \hbar e/c) \,\varepsilon_{klm} B_m
\end{equation}
where we use Einstein's summation convention for repeated indices. From this relation one may derive the commutator $[\pi_k,\vec{S}\cdot\vec{\pi}]=( \rmi \hbar e/c) \,\varepsilon_{klm} S_lB_m$ which in turn leads us to
\begin{equation}\label{pi^2comuteSpi}
  \left[ \vec{\pi}^2,\vec{S}\cdot\vec{\pi}\right] =(e\hbar/c)[\vec{S}\cdot\vec{B},\vec{S}\cdot\vec{\pi}] +
                 (e\hbar/c) S_kS_l\left(\pi_l B_k - B_l \pi_k \right)\,.
\end{equation}
For an arbitrary magnetic field the components of the kinetic momentum do not commute with the components of the magnetic field. However, if we now assume that the magnetic field is constant one may commute in the last term these components. That is, under the assumption that $\vec{B}=const.$ we arrive at
\begin{equation}\label{pi^2comuteSpi2}
  \left[ \vec{\pi}^2,\vec{S}\cdot\vec{\pi}\right] =(2e\hbar/c)[\vec{S}\cdot\vec{B},\vec{S}\cdot\vec{\pi}]\,,
\end{equation}
which in turn results in
\begin{equation}\label{PEcom}
 \left[ \vec{\pi}^2,(\vec{S}\cdot\vec{\pi})^2\right] =\frac{2e\hbar}{c} \left[(\vec{S}\cdot\vec{B}),(\vec{S}\cdot\vec{\pi})^2 \right]\,.
\end{equation}
Note that relation (\ref{pi^2comuteSpi2}) was already given in eq.\ (3.17) of ref.\  \cite{DF1993}.
With the help of (\ref{PEcom}) it is easy to calculate the commutator
\begin{equation}\label{MAs=1app}
  [M_\pm,A]=(g-2)\frac{e\hbar}{2m^2c}\left[(\vec{S}\cdot\vec{B}),(\vec{S}\cdot\vec{\pi})^2 \right] +
  (2g-2)\frac{e\hbar}{2m^2c}\left[\vec{\pi}^2,(\vec{S}\cdot\vec{B}) \right] \,.
\end{equation}
Noting that we have derived this under the assumption of a constant magnetic field the last commutator in above expression vanishes and hence we arrive at eq.\ (\ref{AMS=1}). Note that $[S_k,S_l]=\rmi\varepsilon_{klm}S_m$ and therefore the first term on the right-hand-side above even for a constant magnetic field only vanishes when $g=2$.

With the assumption that the magnetic field is constant and utilising below properties of the spin-one matrices
\begin{equation}\label{Smatices}
  S_iS_jS_k + S_kS_jS_i = \delta_{ij}S_k + \delta_{jk}S_i\,,\qquad \varepsilon_{ijk}S_iS_jB_k =\rmi \vec{S}\cdot\vec{B}
\end{equation}
one may verify the relations (see eq.\ (3.18) and (3.19) in ref.\ \cite{DF1993})
\begin{equation}\label{DFrelations}
\begin{array}{l}
(\vec{S}\cdot\vec{\pi})^4 = \left(\vec{\pi}^2-\frac{2e\hbar}{c}(\vec{S}\cdot\vec{B})\right) (\vec{S}\cdot\vec{\pi})^2 + \frac{e\hbar}{c}(\vec{B}\cdot\vec{\pi})(\vec{S}\cdot\vec{\pi})\,, \\[2mm]
\left\{(\vec{S}\cdot\vec{B}), (\vec{S}\cdot\vec{\pi})^2\right\} = \left(\vec{\pi}^2-\frac{e\hbar}{c}(\vec{S}\cdot\vec{B})\right)(\vec{S}\cdot\vec{B}) + (\vec{B}\cdot\vec{\pi})(\vec{S}\cdot\vec{\pi})\,.
\end{array}
\end{equation}
Noting that for $g=2$ we have
\begin{equation}\label{A_HV_App}
  H_{\rm V} := \frac{\vec{\pi}^2}{2m}-\frac{e\hbar}{mc}(\vec{S}\cdot\vec{B}) \,,\qquad  A = \frac{\vec{\pi}^2}{2m}-\frac{1}{m}(\vec{S}\cdot\vec{\pi})^2
\end{equation}
and with above relations (\ref{DFrelations}) immediately follows that $A^2=H^2_{\rm V}$ as claimed in the main text. Finally, let us mention the explicit form of the energy eigenfunctions can also be found in ref.\ \cite{DF1993}.
\reftitle{References}


\begin{thebibliography}{999}
\bibitem{Klein1926}Klein, O. {Quantentheorie und f\"unfdimensionale Relativit\"atstheorie.} {\em Z.\ Physik} {\bf 1926}, {\em 37}, 895--906.
\bibitem{Gordon1926}Gordon, W. {Der Comptoneffekt nach der Schr\"odingerschen Theorie.} {\em Z.\ Physik} {\bf 1926}, {\em 40}, 117--133.
\bibitem{Dir1928a}Dirac, P.A.M. {The Quantum Theory of the Electron.} {\em Proc.\ Roy.\ Soc.\  A} {\bf 1928}, {\em 117}, 610--624.
\bibitem{Dir1928b}Dirac, P.A.M. {The Quantum Theory of the Electron Part II.} {\em Proc.\ Roy.\ Soc.\  A} {\bf 1928}, {\em 118}, 351--361.
\bibitem{Proca1936}Proca, A. {Sur la th\'{e}orie ondulatoire des électrons positifs et n\'{e}gatifs.} {\em J.\ Phys.\ Radium} {\bf 1936 }, {\em 7}, 347--353.
\bibitem{Dir1936}Dirac, P.A.M. { Relativistic Wave Equations.} {\em  Proc.\ Roy.\ Soc.\  A} {\bf 1936}, {\em 155}, 447--459.
\bibitem{Fierz1939}Fierz, M. {\"Uber die relativistische Theorie kr\"aftefreier Teilchen mit beliebigem Spin.} {\em Helv.\ Phys.\ Acta} {\bf 1939}, {\em 12}, 3--37.
\bibitem{FP1939}Fierz, M.; Pauli, W. { On Relativistic Wave Equations for Particles of Arbitrary Spin in an Electromagnetic Field.} {\em  Proc.\ Roy.\ Soc.\  A} {\bf 1939}, {\em 173}, 211--232.
\bibitem{Bhabha1945}Bhabha, H.J. { Relativistic wave equations for the elementary particles.} {\em Rev.\ Mod.\ Phys.\ } {\bf 1945}, {\em 17}, 200--216.
\bibitem{BW1948}Bargmann, V.; Wigner, E.P. {Group Theoretical Discussion of Relativistic Wave Equations.} {\em Proc.\ Natl.\ Acad.\ Sci.\ U.S.A.\ }{\bf 1948}, {\em 34}, 211--223.
\bibitem{FM1950}Foldy, L.L.; Wouthuysen, S.A. {On the Dirac Theory of Spin 1/2 Particles and Its Non-Relativistic Limit.} {\em Phys.\ Rev.\ }{\bf 1950}, {\em 78}, 29--36.
\bibitem{Foldy1956}Foldy, L.L. {Synthesis of Covariant Particle Equations.} {\em Phys.\ Rev.\ } {\bf 1956}, {\em 102}, 568--581.
\bibitem{FV1958}Feshbach, H.; Villars, F. {Elementary Relativistic Wave Mechanics of Spin 0 and Spin 1/2 Particles.} {\em Rev.\ Mod.\ Phys.\ }{\bf  1958}, {\em 30}, 24--45.
\bibitem{Duffin1938}Duffin, R.J.  {On The Characteristic Matrices of Covariant Systems.} {\em Phys.\ Rev.\ }{\bf 1938}, {\it 54}, 1114; 
\bibitem{Kemmer1939}Kemmer, N. { The particle aspect of meson theory} {\em Proc.\ Roy.\ Soc.\ A} {\bf 1939}, {\it 173}, 91--116;
\bibitem{YST1938}Yukawa H.; Sakata S.; Taketani M. {On the Interaction of Elementary Particles.\ III} {\em Proc.\ Phys.-Math.\ Soc.\ Japan} {\bf 1938}, {\it 20}, 319--340.
\bibitem{ST1940}Sakata, S.; Taketani, M. {On the Wave Equation of Meson} {\em Proc.\ Phys.-Math.\ Soc.\ Japan} {\bf 1940}, {\it 22}, 757--770. 
\bibitem{CS1940}Corben, H.C.; Schwinger J. {The Electromagnetic Properties of Mesotrons } {\em Phys.\ Rev.\ }{\bf 1940}, {\it 58}, 953--968. 
\bibitem{Sch1954}Schr\"odinger, E. {The wave equation for spin 1 in Hamiltonian form.} {\em Proc.\ Roy.\ Soc.\  A} {\bf 1955}, {\em 229}, 39--43.
\bibitem{Silenko2016}Silenko, A. {Exact form of the exponential Foldy-Wouthuysen transformation operator for an arbitrary-spin particle.} {\em Phys.\ Rev.\ A}, {\bf 2016}, {\em 94}, 032104.
\bibitem{Simulik2019}Simulik, V.M. {Relativistic Equations for arbitray Spin, especially for the Spin $s=2$.} {\em Ukr.\ J.\ Phys.\ }{\bf 2019}, {\em 64}, 1064--1068.
\bibitem{Thaller92}Thaller, B. {\it The Dirac Equation}; Springer-Verlag: Berlin, Germany, 1992.
\bibitem{Jun2019}Junker, G. {\it Supersymmetric Methods in Quantum, Statistical and Solid State Physics, Enlarged and revised edition}; IOP Publishing: Bristol, UK, 2019.
\bibitem{Znojil2004}Znoil, M. {Relativistic supersymmetric quantum mechanics based on Klein–Gordon equation} {\em J.\ Phys.\ A} {\bf 2004}, {\it 37}, 9557.
\bibitem{JunIno2018}Junker, G.; Inomata, A. {Path integral and spectral representations for supersymmetric Dirac-Hamiltonians} {\em J.\ Math.\ Phys.} {\bf 2018}, {\it 59}, 052301.
\bibitem{Fock1928}Fock, V. {Bemerkung zur Quantelung des harmonischen Oszillators im Magnetfeld} {\em Zeit.\ f.\ Physik} {\bf 1928}, {\it 47} 446-448.
\bibitem{Landau1930}Landau, L. {Diamagnetismus der Metalle} {\it Zeit.\ f.\ Physik} {\bf 1930}, {\it 64}, 629-637; 
\bibitem{AC78}Aharonov, Y.; Casher, A. {Ground state of a spin-$\frac{1}{2}$ charged particle in a two-dimensional magnetic field} {\em Phys.\ Rev. D} {\bf 1978}, {\bf 19}, 2461-2462;
\bibitem{Rabi1928}Rabi, I.I. {Das freie Elektron im homogenen Magnetfeld nach der Diracschen Theorie} {\em Zeit.\ f.\ Physik} {\bf 1928}, {\it 49}, 507–511; 
\bibitem{YB1963}Young, J.A.; Bludman, S.A.  {Electromagnetic Properties of a Charged Vector Meson} {\it Phys.\ Rev.\ }{\bf 1963}, {\it 131}, 2326-2334;
\bibitem{KLG1971}Krase, L.D.; Lu, P.; Good Jr, R.H. {Stationary States of a Spin-1 Particle in a Constant Magnetic Field} {\it  Phys.\ Rev.\ D} {\bf 1971}, {\it 3}, 1275--12798;
\bibitem{TY1971}Tsai, W.-y.; Yildiz, A. {Motion of Charged Particles in a Homogeneous Magnetic Field} {\it Phys.\ Rev.\ D} {\bf 1971} {\it 4} 3643--3648;
\bibitem{GT1971}Goldman, T.; Tsai, W.-y. {Motion of Charged Particles in a Homogeneous Magnetic Field II} {\it Phys.\ Rev.\ D} {\bf 1971}, {\bf 4}, 3648--3651;
\bibitem{Weaver1976}Weaver, D.L. {Application of a single-particle-theory calculation of $g-2$ to spin one} {\it Phys.\ Rev.\ D} {\bf 1976}, {\it 14}, 2824--2825;
\bibitem{DF1993}Daicic, J.; Frankel, N.E. {Relativistic spin-1 bosons in a magnetic field} {\it J.\ Phys.\ A: Math. Gen.\ }{\bf 1993}, {\it 26}, 1397--1408;
\bibitem{Sil2014}Silenko, A.J.  {High precision description and new properties of a spin-1 particle in a magnetic field} {\it Phys.\ Rev.\ D} {\bf 2014}, {\it 89}, 121701(R);
\bibitem{Jun2020}Junker, G. {Supersymmetric quantum mechanics requires $g = 2$ for vector bosons} {\it Eur.\ Phys.\ Lett.\ }{\bf 2020}, {\it 130} 30003.
\bibitem{Guertin1974}Guertin, R.F. {Relativistic Hamiltonian Equations for Any Spin} {\it Ann.\ Phys.\ }{\bf 1974}, {\it 88}, 504--553.
\end{thebibliography}
\end{document}